\documentstyle[11pt,paspconf,psfig]{article}

\def\etal{et al.~}

\def\lb{line blanketing}

\def\apj{ApJ }
\def\apjs{ApJS }

\def\aj{AJ }
\def\aa{A\&A }
\def\aas{A\&AS }
\def\araa{ARA\&A }
\def\mnras{MNRAS }

\def\pasp{PASP }

\def\mdot{\ifmmode  \dot{M} \else $\dot{M}$\fi}
\def\msun{\ifmmode M_{\odot} \else $M_{\odot}$\fi}
\def\vinf{\ifmmode v_{\infty} \else $v_{\infty}$\fi}
\def\teff{\ifmmode T_{\rm eff} \else $T_{\rm eff}$\fi}
\def\logg{\ifmmode \log g \else $\log g$\fi}
\def\loggeff{\ifmmode \log g_{\rm eff} \else $\log g_{\rm eff}$\fi}
\def\rstar{\ifmmode R_{\star} \else $R_{\star}$\fi}
\def\lstar{\ifmmode L_{\star} \else $L_{\star}$\fi}
\def\mstar{\ifmmode M_{\star} \else $M_{\star}$\fi}
\def\rsun{\ifmmode R_{\odot} \else $R_{\odot}$\fi}
\def\lsun{\ifmmode L_{\odot} \else $L_{\odot}$\fi}

\def\hi{H~{\sc i}}
\def\hii{H~{\sc ii}}
\def\hei{He~{\sc i}}
\def\heii{He~{\sc ii}}

\def\he0{\ifmmode {\rm He^{\circ}} \else $\rm He^{\circ}$\fi}
\def\hep{\ifmmode {\rm He^+} \else $\rm He^{+}$\fi}
\def\hepp{\ifmmode {\rm He^{2+}} \else $\rm He^{2+}$\fi}
\def\halpha{\ifmmode {\rm H{\alpha}} \else $\rm H{\alpha}$\fi}
\def\hgamma{\ifmmode {\rm H{\gamma}} \else $\rm H{\gamma}$\fi}

\begin{document}

\title{Massive Star Evolution in Different Environments\footnotemark[1]} 
\footnotetext[1]{Invited review to appear in ``Active Galactic Nuclei, 
Dense Stellar Systems, and Galactic Environments'', Eds. A. Baker, S.A. Lamb, 
J.J. Perry, ASP Conf.~Series, 1998}
\author{Daniel Schaerer $^{1,2,3}$}
\affil{$^1$ Laboratoire d'Astrophysique, Observatoire Midi-Pyr\'en\'ees,
	14, Av. E. Belin, F-31400 Toulouse, France (schaerer@obs-mip.fr)}
\affil{$^2$ Space Telescope Science Institute, 
	Baltimore, MD 21218, USA}
\affil{$^3$ Geneva Observatory, CH-1290 Sauverny, Switzerland}

\begin{abstract}
We review the properties of massive star evolution in different
environments, where the major environmental factor is metallicity.
Comparisons between evolutionary models and observations
of massive OB, WR stars and related objects are presented. 
We also review several observations asking for future improvements of 
stellar models and theoretical developments in this respect.
We summarize evolutionary scenarios for the most massive stars and 
try to clarify recent questions regarding their evolutionary status
as core-H or core-He burning objects.

Another environmental effect, which might affect stellar evolution
is a cluster environment with a high stellar density.
As test cases of massive star evolution in dense clusters
we summarize recent work on the densest known resolved young 
clusters: R136, NGC 3603, and the three Galactic Center (GC) star 
clusters -- the central cluster, Quintuplet and the ``Arches'' cluster.
For the central cluster we present new comparisons between stellar parameters 
of emission line stars derived by Najarro \etal (1994, 1997), 
and appropriate evolutionary models.
From their parameters we argue that most of these stars can be regarded 
as WNL stars, and do hence not necessarily represent a peculiar class. 
We suggest that some apparent differences with well known WR stars 
can be understood in terms of their core burning stage and/or
other changes due to a high metallicity.
Based on our present knowledge 
we conclude that in young clusters with central stellar densities up to 
$\rho_c \sim 10^{5-6} \, \msun {\rm pc}^{-3}$
no compelling evidence for a secondary effect influencing the evolution 
of massive stars has yet been found.

\end{abstract}


\vspace*{-1cm}
\section{Introduction}
Massive stars play an important role in driving the evolution of
galaxies. Through their strong radiation field and their stellar winds 
O type stars and their evolved descendents, the Wolf--Rayet (WR) stars, 
are major contributors in UV radiation, mass, momentum, and mechanical 
energy input to the interstellar medium (ISM).
They are thus an important source to ionize the ISM and power the
far-infrared luminosities through the heating of dust.
As progenitors of supernovae, massive stars are agents of nucleosynthesis,
and they may provide strong feedback mechanisms acting on new star
formation processes.
Therefore massive star evolution is a key study in the exploration
of many facets of the Universe.

While for obvious reasons all the above topics have ``traditionally''
been studied in our Galaxy or nearby objects, observational progress 
now allows us to distinguish similar processes up to very large 
distances and at different scales in a large variety of objects including 
extragalactic \hii\ regions, IRAS galaxies, and various kinds of emission 
line galaxies such as starbursts and also Active Galactic Nuclei (AGN).
In many of those direct or indirect signatures reveal the presence 
of massive stars.
For the understanding of these remote and/or complex systems a good 
knowledge of its basic constituents (stars \& gas) and the fundamental
processes governing and linking them is prerequisite.

In the present review we will mainly concentrate on massive star 
evolution in different environments. 
In most astrophysical conditions the major ``environmental'' effect
on stellar evolution is the initial composition (metallicity) of the 
star.
In more general situations one can, however, have a multitude of 
mechanical and radiative effects which may influence the evolution 
of stars: 
star--star interactions (collisions, captures etc.), 
star--ISM interactions in a very dense interstellar medium,
star--accretion disk interactions in AGN,
external gravitation fields, external radiation fields etc.
Such effects may indeed be of particular importance for studies of 
very dense stellar systems, AGN and related phenomena discussed
in this volume (see e.g.~the contributions of Murphy, Perry, Baker).

Our aim here is rather conservative in this respect. We first try to 
understand stellar evolution in simple systems (field and clusters in 
Local Group galaxies) and study the effects of the ``environmental'' 
effects which are the most important in these cases.
It is the hope that investigations on stellar evolution in more 
``extreme'' conditions will benefit from this understanding.

In Sect.~2 we briefly summarize IMF determinations in young 
clusters. We then review the physics and properties of stellar models 
for massive stars in Sect.~3. This section builds on the extensive 
review of Maeder \& Conti (1994), and discusses more recent work in this field.
In Sect.~4 we summarize comparisons between observations in dense 
young clusters and stellar evolution models.
Section 5 is devoted to a detailed discussion about the massive
stars in the Galactic Center star cluster.

\section{Massive star census and IMF determinations}
\label{s_imf}
Stellar counts in associations are the most direct way to obtain
estimates of the slope of the initial mass function (IMF).
The pioneering work of Massey and coworkers using both
photometry and spectroscopy provides a homogeneous
approach to determine the IMF in different environments. 

From analysis of nearly twenty Galactic and LMC associations
they find no statistically significant variation in IMF slopes and an
average value of $\Gamma = -1.0 \pm 0.1$ for stars with masses
$M > 7 \msun$\footnote{The Salpeter value is $\Gamma$ = -1.35 in this 
notation.} (Massey \etal 1995b, cf.~also compilation in MC94).
No differences are found between Galactic and Magellanic OB associations
indicating that massive star formation in clusters proceeds independently
of metallicity, at least over the range considered in their work.
This finding is in contrast to the prevailing view (Shields \& Tinsley
1976) that $\Gamma$ becomes steeper with increasing abundance.
In addition Massey \etal (1995b) do not find metallicity variations 
of the upper mass limit, which can reach very large values ($M_{\rm up}
\sim$ 100 - 120 \msun).
Regarding the lower end of the mass function recent ground-based and 
HST observations of R136 and NGC 3603 (Hofmann \etal 1995, Hunter \etal 
1995, Brandl \etal 1996, Eisenhauer \etal 1998) do not show any lack 
of low-mass stars. 
This supports the findings of Satyapal \etal (1996) who question the 
claims of a truncated IMF in starburst galaxies (e.g.~Rieke \etal 1993).
Recent reviews about the IMF can be found in the volume of Gilmore \etal
(1998).


Results concerning the IMF in some dense young clusters will be briefly
mentioned in Sect.~\ref{s_clusters}

\section{Stellar evolution models and observations}
\label{s_evol}
\subsection{Input physics}
A large number of grids of stellar models at different metallicities $Z$
based on various physical assumptions have been published in the
recent years (see MC94 and Maeder 1996 for a compilation).
The most critical ingredients for models of massive stars discussed
here are the mass loss prescription and the treatment of convection and
mixing in the stellar interiors.

Since evaporation by stellar winds is dominant for stars with
initial masses $M_{\rm ini} \ga 20 \msun$ (Maeder 1991a) all model 
predictions are influenced by the adopted mass loss rates \mdot. Given
the present discrepancies between observed values of \mdot\ and
predictions from the radiation driven wind theory (cf.~Lamers \&
Leitherer 1993, Puls \etal 1996, Schaerer \etal 1996a) empirical values
are presently used. 
Most authors use those from the compilation of de Jager \etal (1988),
while the newly derived empirical formulae of Lamers \& Cassinelli (1996) 
have not been applied so far.
The average mass loss rates possibly being too low (Schaerer \& Maeder 1992),
Meynet \etal (1994) adopted higher values, which allows a good agreement
for numerous observational properties (see Maeder \& Meynet 1994).
Additional mixing processes (see below) might, however, also mimic
a similar behaviour and hence allow to reduce the required mass loss rates.

The treatment of convection and mixing is a major uncertainty in massive 
stars models. Schematically the following assumptions can be identified:
Schwarzschild or Ledoux criterion, core overshooting, overshooting
below the convective envelope, semiconvection or semiconvective diffusion,
turbulent diffusion or other forms of rotational mixing.
A critical discussion of the importance of these processes can be found
in MC94. 
In view of the large number of evidence pointing towards a need of additional 
mixing processes (see MC94, Maeder 1995c, and below) a more unified physical
description of these processes seems required.
After earlier works on this subject (e.g.~Maeder 1987, Langer 1992)
first progress in this direction has been made recently (Maeder 1995a, 
Fliegner \etal 1996, Meynet \& Maeder 1997, Maeder \& Zahn 1998).

In addition to the usual ingredients, WR models require specific attention
on a number of points (mass loss rates, equation of state etc.). 
These are discussed in MC94.

\subsection{Metallicity Effects in Massive Stars}
Metallicity is a key factor which influences the evolution and hence the 
populations of massive stars.
To be able to distinguish this effect from possible variations of 
star formation rates, changes of the IMF etc., one must consider
the following dependences on the metallicity $Z$ (cf.~MC94):

{\em 1) Nuclear production:} Compositions changes may influence the
nuclear reactions. For example, in massive stars a low $Z$ can produce
a more active H-burning shell, which favours a blue location during
part or the entire He-burning phase.

{\em 2) Opacity effects:} Since electron scattering, which is independent of $Z$,
is the dominant opacity source in the interior of massive stars metallicity has
no important direct effect on their inner structure.

{\em 3) Stellar wind:} However, in the external layers where bound-free and 
bound-bound opacities become important, $Z$ has a strong influence on the opacity
and hence on the atmospheres and winds. 
The metallicity dependence of the mass loss rates (e.g.~Kudritzki \etal
1987) is the main effect by which $Z$ influences the evolution of massive stars 
(Maeder 1991a).

{\em 4) He content:} An increasing He content with metallicity
as established from low-$Z$ \hii\ regions (e.g.~Pagel \etal 1992)
has a direct effect on the models (different fuel reservoir, different
interior opacity).

Before summarizing some of the main properties of massive star
models at various metallicities we shall first review recent work on their
pre-main sequence. Properties related to WR stars will be 
discussed later (see Sect.~\ref{s_wr}).

\subsection{Evolution up to the End of the Main Sequence }

\subsubsection{Pre-Main Sequence Evolution:}
Generally the pre-MS evolution of massive stars is considered to be 
very short (typically 1 \% of the MS lifetime) and is 
therefore neglected in most studies. Recent observational 
(see e.g.~Churchwell 1993, Hanson \& Conti 1995) and theoretical 
progress (e.g.~Yorke 1993, Hollenbach \etal 1994, Beech \& Mitalas 1994, 
Jijina \& Adams 1996, Bernasconi \& Maeder 1996, Bonnel \etal 1998) 
have lead to a considerably distinct picture of the early evolution of 
massive stars. Here we will briefly summarize the work of Bernasconi 
\& Maeder (1996).

Following the accretion scenario first proposed by Palla \& Stahler
(1990) these authors calculate the evolution of an initially 0.8 \msun\
protostellar core until the total mass has reached typical values
for high mass stars ($\sim$ 60 -- 100 \msun). A basic parameter is 
the (variable) mass accretion rate, which Bernasconi \& Maeder derive 
from the equilibrium equations of cloud models, accounting both for 
thermal pressure and a non-thermal support.
The major results from their work are the following:
{\em 1)} The accretion phase for massive stars lasts some 2-2.5 Myr, and is
therefore nearly comparable to the usual MS lifetimes.
{\em 2)} Newly formed massive stars with $M \ga 40 \msun$ may have already
burnt a substantial fraction of their central hydrogen and hence have evolved 
away from the classical zero-age MS (ZAMS) at the time they emerge from their 
parental cloud.
As a consequence their remaining MS lifetimes are correspondingly reduced.
{\em 3)} Higher turbulence in the molecular cloud leads to larger accretion
rates. This in turn implies that stars of higher mass can be formed.

The second point may explain an apparent lack of O-type stars close to the
formal ZAMS (Garmany \etal 1982) although this observational finding is
not very well established (see Massey \etal 1995a).
The last point of the scenario of Bernasconi \& Maeder (1996) in particular
provides interesting links between the properties of the environment and the 
formation and evolution of massive stars.
Many implications remain to be worked out and the models have to be confronted 
to observations.
A more consistent picture also explaining the role of ultra-compact
\hii\ regions (see e.g.~Churchwell 1993) in the framework of massive star 
formation would be highly desirable. 

\subsubsection{Main Sequence Evolution:}
The position of the tracks in the HR-diagram and the lifetimes in the
various evolutionary phases change with the metallicity $Z$.
For massive stars at low $Z$ the formal ZAMS is shifted to the blue and 
the luminosity is slightly lower for a given mass.
Due to their lower luminosity and the larger initial H content, massive stars 
have longer H-burning lifetimes $t_H$ at low $Z$ (typical differences 
between $Z_\odot/20$ and $2 Z_\odot$ are 35 \% for a 20 \msun\ star).
However, lifetimes in the He-burning phase are generally shorter at low
$Z$ due to the lower mass loss rates which lead to higher luminosities
in these phases.
$t_{\rm H}/t_{\rm He}$ ranges from typically 9--10 \% at $Z=0.001$ to
11--19 \% at $Z=0.04$ for $M \ge 15 \msun$. Adopting the large mass loss 
rate of Meynet \etal (1994) can lead to $t_{\rm H}/t_{\rm He}$ up to 
$\sim 0.5$ at high $Z$.
For more details we refer to the review of MC94 and references therein. 

Surface abundances of He and CNO products represent extremely important
tests of stellar evolution. Evidence for CN processing is available by
He and N enhancements together with C depletion, while O is only gradually
depleted in advanced stages of processing. The observations of Herrero \etal
(1992) and Gies \& Lambert (1992) show that most MS OB stars have normal
He and N abundances.
These elements are, however, enriched in fast rotators (Herrero \etal 
1992), which suggests some additional mixing process related to
rotation.

\subsection{Post-MS evolution and Supergiants:}
Very fundamental properties, like the relative lifetimes spent in the
H and He-burning phases $t_{\rm H}/t_{\rm He}$, are quite well established. 
For example the comparison of Meynet (1993) with open clusters provides 
strong constraints on masses $M \la 20 \msun$.
More subtle properties turn out to be the location of the He-burning stars 
in the HR-diagram and their surface abundances.

Indeed many problems and uncertainties remain about
supergiants, for which evolution is more uncertain than for WR stars.
The reason is that WR stars are dominated by powerful mass loss 
(``evaporation'') which overwhelms most effects related to uncertainties
in convection and mixing. Supergiants are often close to a neutral state
between blue and red in the HR-diagram where even small changes in convection
and mixing processes can considerably alter their evolution.
Let us now briefly summarize the major difficulties arising for supergiants
(see also MC94).

{\em 1) Surface abundances: } 
The basic result is still that by Walborn (1976, 1988)
who showed that ordinary OB supergiants have He and N enrichment as a result
of CNO processing, while only the small group of peculiar OBC stars have normal
cosmic abundance ratios. His results are confirmed by many recent studies
including Galactic OB stars (Howarth \& Prinja 1989, Herrero \etal 1992, 
Gies \& Lambert 1992), LMC stars (e.g.~Lennon \etal 1991, Fitzpatrick \&
Bohannan 1993), and SMC B-type supergiants (Lennon \etal 1991, Lennon 1997).
Thus He and N enrichment appears to be the general rule among B-supergiants,
which places strong constraints on stellar evolution models.

The most simple explanation for the He and N enrichment is that the blue
supergiants are on the blue loops after a first a red supergiant phase,
where they have experienced dredge-up modifying the surface abundances.
Difficulties of this scenario are, however, that current models do not 
necessarily predict blue loops of the ``right'' extension and at the 
required luminosities. More importantly the enrichment predicted by the
1st dredge-up does not seem to be high enough to account for the observations,
as e.g.~shown by Venn (1993, 1995).
Other explanations face similar difficulties.
Additional mixing processes seem thus to be required in massive stars
(see Maeder 1995c).

{\em 2) Distribution of supergiants in the HR-diagram:} 
There appear to be many more stars outside the MS band than predicted
(see e.g.~Stothers \& Chin 1977, Meylan \& Maeder 1982).
The so-called blue Hertzsprung gap predicted by most stellar models to occur at 
the end of the MS does not seem to be observed. A summary of possible
solutions can be found in MC94. 
At present the question is not settled.

{\em 3) Blue/red ratio of supergiants in galaxies:}
The observed number ratio B/R of blue to red supergiants increases
with metallicity $Z$ by typically a factor of 10 from the SMC to the
solar neighbourhood (Humphreys \& McElroy 1984, Langer \& Maeder 1995).
As mentioned earlier, the blue or red location is extremely sensitive
to different model assumptions.
Although generally a given set of stellar models has no difficulty
of explaining the observed B/R at a given $Z$ (e.g.~through small
changes of mass loss) all stellar models so far predict the 
opposite behaviour of B/R with metallicity (Langer \& Maeder) !
These authors suggest a connection of the B/R problem with internal
mixing.

\subsection{Basic properties and evolution of WR stars:}
\label{s_wr}
WR stars are generally considered to be bare cores which have mainly
evolved from initially massive stars ($M_{\rm ini} \ga$ 25--40 \msun).
Close binaries may also lose their outer layers from Roche lobe overflow.
[For detailed reviews on WR stars see e.g.~MC94 and references therein, 
or the recent Li\`ege proceedings (Vreux \etal 1996)].
WR stars form the following consistent chemical sequence 
(E stands for early, L for late):

\begin{itemize}
\item[]{\em WNL:} products of the CNO cycle at equilibrium with (in general)
	H present
\item[]{\em WNE:} CNO equilibrium products with no H
\item[]{\em WN/WC:} Rare group ($\sim$ 4 \%), where products of both the CNO cycle 
	and the 3$\alpha$ reaction are present (transition case, see 
	e.g.~Langer 1991, Crowther \etal 1995b)
\item[]{\em WCL:} He-burning products (He, C, O) present with low values of
	 (C+O)/He
\item[]{\em WCE:} same but with high (C+O)/He
\item[]{\em WO:} same but with O $\ge$ C
\end{itemize}

It is important to note that this sequence is not fully described by all
stars. What phases a star actually evolves through and the duration in thoses
phases is strongly dependent on its initial mass and metallicity (cf.~below).

\subsubsection{Evolutionary scenarios:}
From recent work the following filiation scheme leading to a final SN 
explosion (e.g.~Woosley \etal 1993, 1995) can be drawn for massive
stars (cf.~Maeder 1991b, 1996a, 1997, Langer \etal 1994, Crowther \etal 1995a, 
Crowther \& Smith 1997, Pasquali \etal 1997):

\begin{itemize}
\item[] {\bf $M \ga$ 60 \msun:} \newline
	O --- Of --- WNL+abs --- WN7 --- (WNE) --- WCL --- WCE -- SN \newline
	\hspace*{5.4cm} At low $Z$: ... WN7 --- WCE -- SN
\item[] {\bf $M \simeq$ 40 -- 60 \msun:} \newline
	O --- Of --- LBV + Ofpe/WN9 --- WN8 --- WNE --- WCE -- SN
\item[] {\bf $M \simeq$ 25 -- 40 \msun:} \newline
	O ---  (BSG) --- RSG --- (BSG) --- WNE --- (WCE) -- SN
\item[] {\bf $M \la$ 25 \msun:} \newline
	O --- (BSG) --- RSG --- BSG --- RSG  -- SN \newline
	\hspace*{4cm} 		YSG  -- SN
\end{itemize}
Parenthesis indicate uncertain or very short phases. 
``LBV + Ofpe/WN9'' stands for an intermediate LBV phase, with Ofpe/WN9
stars (or equivalently WN9-11 according to Crowther \& Smith 1997)
representing a hot dormant LBV phase (see Nota \etal 1996, Crowther \& Smith).
According to different galactic locations (reflecting different metallicities)
one has the following differences in the WR subtype evolution for $M \ga$ 40
\msun. Inner regions: WNL $\rightarrow$ WCL, 
outer regions: WNL $\rightarrow$ WCE $\rightarrow$ WO.

Although the evolutionary paths are quite well understood on the whole,
several uncertainties and open questions remain. These will be
briefly discussed in the following:

\noindent {\em Evolution through LBV phase ?}
For $M \ga$ 60 \msun\ it is still not completely clear whether the most
massive stars go through the LBV phases (cf.~Schaller \etal 1992, Langer
\etal 1994, Pasquali \etal 1997) or whether they avoid this phase 
(Crowther \etal 1995a, cf.~also Meynet \etal 1994).
LBV stars and their relations to other classes are discussed in the
volume of Nota \& Lamers (1997).

\noindent {\em Are WNL and Ofpe/WN stars core H or He burning objects ?}
The previous point is also related to the question whether some WNL stars 
are in the core-H burning phase as already suggested early by Conti (1976).
Such a scenario is indeed a natural 
outcome for the most massive stars given their strong mass loss (e.g.~Schaller 
\etal 1992, Meynet \etal 1994)
--- no assumption is made about the burning source of WR stars in such models,
and {\em a priori} H/He surface abundances do not allow to determine the core 
burning source.
In fact the models of Meynet \etal
e.g.~predict that at $Z=0.02$ all WNL stars\footnote{The WNL phase in these
models is defined by $\log \teff >4.$ and a hydrogen abundance $0 < X \le 0.4$
in mass fraction.} with $\log L/\lsun \ga 6$ should
still be core-H burning objects. This limit decreases at higher $Z$,
which might in particular help to understand WR-like stars in the Galactic
Center (see Sect.~\ref{s_gc})\footnote{Additional mixing processes may also
ease the formation of core H-burning WR stars}.

Observational evidence in favour of such a scenario comes from similarities
of some Of and WN spectra (Conti \etal 1995, Morris \etal 1996), analysis of 
the young cluster NGC 3603 and the most luminous stars in R136a (Drissen \etal 
1995, de Koter \etal 1997, Crowther \& Dessart 1998, see Sect.~\ref{s_clusters}),
and the very high mass of a WN7 star recently measured by Rauw \etal (1996).
If this is the case, the most massive O stars might well evolve directly 
to WNL stars and avoid the LBV phase.

An alternative evolutionary scenario including a phase of strong mass loss 
on the main sequence due to pulsational instabilities was proposed by 
Langer \etal (1994).
This scenario predicts two distinct WN phases separated by an LBV phase,
the first WN phase occurring during core H-burning. 
Although it describes well H-rich WN stars, P Cygni type LBVs (Langer \etal 
1994) and might be supported by observations in R136 (Heap \etal 1994, but
cf.~de Koter \etal 1997), 
too few WN stars with low H abundance seem to be predicted (Maeder 1995b).

In the line of the Langer \etal scenario, Pasquali \etal (1997) have recently 
suggested the evolutionary sequence: O -- Of -- H-rich WNL -- Ofpe/WN9, and 
argue that even the less massive (LMC) Ofpe/WN9 stars must be core H-burning 
objects.
This contrasts the scenario of Crowther \& Smith (1997) summarized above.
We also note that in view of remaining differences in evolutionary 
models a different conclusion about the core burning source of the analysed
Ofpe/WN9 stars is very well possible. Their claimed constraint from
surface temperatures and abundances is not conclusive (see e.g.~the models
of Meynet et al.).

\subsubsection{Properties of WR stars:}
The main properties of WR star models have been amply discussed by 
Maeder \& Meynet (1994).
Through the key relations for WR stars (mass-luminosity relation: Maeder 1983,
mass-\mdot\ relation: Langer 1989) all properties of WNE and WC
stars are related (Schaerer \& Maeder 1992).

The behaviour of (C+O)/He in WC/WO stars for different metallicities $Z$
is of particular interest: At high $Z$ (high mass loss in pre-WR phases)
the He-burning core is revealed earlier in the evolution and shows thus
low (C+O)/He ratios, while at low $Z$ the He-burning core is revealed much
later, i.e.~with very high (C+O)/He (see e.g.~Maeder \& Meynet 1994).
This explains nicely the following main observed properties of WC stars:
{\em 1)} If WC stars exist at low $Z$ they are of types WCE and WO,
{\em 2)} WCL stars exist only at high $Z$,
{\em 3)} At a given $Z$, the luminosities of WCL stars are higher than those
	of WCE (cf.~Smith \& Maeder 1991), and
{\em 4)} For a given WC subtype the luminosities are higher at lower $Z$
	(cf.~Kingsburgh \etal 1995).

Of major importance for the understanding of WR populations is the 
behaviour of the lifetimes $t_{\rm WR}$ of WR stars as a function of initial 
mass and metallicity. The most important effect is a strong increase of 
$t_{\rm WR}$ with initial mass and $Z$ (due to increased mass loss), and
an increase of the threshold mass for WR formation from single stars
(see Maeder \& Meynet 1994).
These effects are responsible for the observed increase of WR/O ratios 
in nearby galaxies (cf.~MC94). The general increase of the WC/WN number 
with $Z$ is also well accounted for. 
Further comparisons are found in Maeder \& Meynet (1994).

\section{Massive Star Evolution in Clusters}
\label{s_clusters}
Studying stellar clusters is of fundamental importance for our
understanding of stellar evolution. Observations in clusters, which
are thought to be formed coevally, provide the most stringent constraints 
on evolutionary models.

A large number of galactic open clusters with ages from 4 Myr to 9.5 Gyr was 
analysed by Meynet \etal (1993).
Of particular interest for the evolution of massive stars is the recent work 
of Massey \etal (1995a) on OB associations in the LMC and SMC.
Both studies find that current evolutionary models show a good agreement 
with observed colour-magnitude diagrams and the stellar distribution in the 
HR-diagram. Stronger tests could be obtained from additionally considering
chemical surface abundances in these cluster/associations.

An important aim of the present work is to discuss and test massive star 
evolution in a particular environment, namely in dense stellar clusters. 
As discussed earlier these are of extreme interest for the understanding
of starburst clusters, giant extragalactic \hii\ regions, the Galactic
Center (see next Section), and possibly also stellar systems in AGN.
We restrict our study to the densest objects where the massive star 
population can still be resolved and hence stars can be analysed
individually.

\subsection{R136 in 30 Doradus:}
The massive star cluster R136 is often considered as a Rosetta Stone
for starbursts (see e.g.~Walborn 1991). It has been extensively observed
and its stellar content has well been resolved by HST and ground-based
observations (e.g.~Campbell \etal 1992, Pehlemann \etal 1992, 
Malumuth \& Heap 1994, Hunter \etal 1995, Brandl \etal 1996, Massey 
\& Hunter 1998).
Its central density is estimated to be 
$\rho_c \sim 10^{4-5} \, \msun \, {\rm pc}^{-3}$
(Hunter et al., Brandl et al.).
Some authors have found a possible flattening of the IMF towards the 
cluster center (Malumuth \& Heap 1994, Brandl \etal 1996; but cf.\
~Hunter \etal 1995, Massey \& Hunter 1998).


From several methods (optical and IR photometry, UV and optical spectroscopy)
most studies derive a cluster age of $\sim$ 1--5 Myr with a small age spread
(Hunter \etal 1995, Brandl \etal 1996, de Koter 1998, Massey \& Hunter 1998);
the numerous most massive O3 stars and associated objects may well be very young
($\sim$ 1--2 Myr, de Koter \etal 1997, Massey \& Hunter).

So far these studies show that the stellar population from $\sim$ 2.8 \msun\
up to the most massive OB and WR stars can well be explained by standard 
evolutionary models appropriate to the metallicity of 30 Dor.
Stringent constraints on massive star evolution is obtained from modeling
observations of individual stars. Such analysis have recently been 
possible in the core of R136a using GHRS spectra (Heap \etal 1991, 1994,
de Koter \etal 1994, Pauldrach \etal 1994). Including the most recent study 
of de Koter \etal (1997) and Crowther \& Dessart (1998), four stars classified 
as O3f/WN and WN have now been analysed quantitatively.
De Koter \etal (1997) and Crowther \& Dessart find that due to their huge 
luminosities and very 
strong mass loss, some of these objects have WR like spectral appearance despite 
appearing to be relatively H rich.
The properties of these objects are found to be in good agreement with 
predictions for young massive core H-burning stars from Meynet \etal (1994). 
Interestingly the observed mass loss rates of the most luminous objects 
(de Koter \etal 1997) seem to be even slightly higher than the high values 
adopted by Meynet et al.
Stronger constraints on surface abundances of members of the R136 cluster
would be very useful.

\subsection{The Galactic starburst NGC 3603:}
The central stellar mass density in the Galactic giant \hii\ region NGC 3603
($\rho_c \sim 2.\, 10^5 \, \msun \, {\rm pc}^{-3}$, Hofmann \etal 1995) 
is comparable to or even exceeds that of R136.
An early comparison between these two objects is found in Moffat \etal (1985).
Recent high spatial resolution observations analysing the stellar population, 
IMF, and related properties have been presented by Moffat \etal (1994), 
Hofmann \etal (1995) and Eisenhauer \etal (1998).
The stellar population and the age ($\sim$ 3 Myr) of NGC 3603 is very 
similar to R136. IMF slopes (for the mass range $\sim$ 15--60 \msun) of 
$\Gamma=-1.4 \pm 0.6$ and $\Gamma=-1.59 \pm 0.22$ have been derived
by Moffat \etal and Hofmann \etal respectively. 
A discussion of the low mass population, the lower end of the IMF and 
comparisons with other star-forming regions is given in Eisenhauer \etal
(1998).

The photometric study of Hofmann \etal (1995) shows that the 
individual stars (down to $\sim$ 15 \msun) can well be described by 
standard evolutionary models.
We note that for an age this young, the recent Meynet \etal tracks can 
also naturally explain the simultaneous presence of O3 and 
WNL stars observed by Drissen \etal (1995). Some of the WN stars are then
indeed expected to be in the H-burning phase as suggested by these 
authors.
At the present state we therefore do not see compelling evidence 
requiring to invoke an enhancement of binaries due to the large
stellar density as suggested by Tamblyn (1996).
A quantitative analysis of three WR stars in the core of NGC 3603 was
presented recently by Crowther \& Dessart (1998).

\section{Star clusters in the Galactic Center}
\label{s_gc}
Given its proximity and the possible presence of a central black hole,
the center of our Galaxy deserves a particular interest for studies of 
nuclear activity in galaxies.
Furthermore the extreme density (stellar densities of typically 
$\rho_\star \ga 2. \, 10^6 \, \msun \, {\rm pc}^{-3}$ for $r < 0.5$ pc,
Krabbe \etal 1995) and the presence of massive stars single out the 
central star cluster as the best laboratory to study massive star
evolution in an extreme environment.

A general introduction to the Galactic Center (GC) and ample discussion 
about related subjects can be found in the excellent reviews of 
Genzel \etal (1994) and Morris \& Serabyn (1996).
More recent work of Genzel \etal (1996, 1997) and Ozernoy \& Genzel (1996)
present new dynamical studies and address the question of accretion onto
the putative black hole respectively. 
In the context of the present contribution we will mostly concentrate on 
the emission line stars present in the central cluster (i.e.\ within
$\sim$ 1 pc of Sgr A*). 
The numerous late type stars and interesting questions related to their 
population and dynamics are e.g.~discussed in Genzel \etal (1994, 1996), 
Blum \etal (1996) and references therein. A recent adaptive optics 
high angular resolution study of the stellar content near the GC is given
in Davidge \etal (1997).
The other major young clusters located near the GC are briefly discussed
in Sect.\ \ref{s_arches}.

\subsection{The central star cluster}
Since the discovery by Forrest \etal (1987) and Allen \etal (1990) of an 
unusual star (the now so-called AF star) close to Sgr A$^\star$
with broad near-IR \hei/\hi\ emission lines numerous similar sources have
been discovered (Krabbe \etal 1991, Blum \etal 1995a, Krabbe \etal 1995,
Tamblyn \etal 1996).  
Their IR luminosity, near-IR colours, and little or no CO absorption in their
spectra indicate that they may be early-type stars mass losing stars.
Based on their $K$-band spectra most of these objects were identified with
Ofpe/WN9 stars (e.g.~Krabbe \etal 1995) although important differences 
in equivalent widths and velocity widths exist (e.g.~Blum \etal 1995b).
These stars, also called ``slash stars''
(Walborn 1982, Bohannan \& Walborn 1989), represent a rare class of objects, 
which are thought to be in an intermediate (henceforth short) evolutionary 
phase between massive main sequence Of stars and Wolf-Rayet (WR) stars
(see Sect.~\ref{s_wr}).

The presence of young massive stars in a cluster around the GC 
clearly indicates an important recent activity of star formation.
At first sight, the large number of objects with fairly uncommon spectral
appearance is, however, very surprising.
The fundamental questions regarding the emission line sources are thus:
\begin{itemize}
\item[{\em 1)}] {\em Are these objects recently formed massive stars ?}
Alternatively, and in order to circumvent difficulties related to the
``hostile'' conditions of star formation in the GC, Morris (1993) proposed 
that the \hei\ stars are 10 \msun\ black holes that have collided with giants. 
In this process they would have acquired 
a dense helium rich atmosphere which reprocesses the luminosity of the underlying
slowly accreting black hole, and might mimic blue \hei\ stars with mass loss.

\item[{\em 2)}] If {\em 1)} is true: {\em Did these massive stars form (2a) 
and evolve (2b) ``normally'' ?}
It might well be that 2a and 2b cannot be answered affirmatively.
Indeed, if the very high stellar densities inferred from the number density
distributions of 2 $\mu$m sources (Krabbe \etal 1995) are correct, collisions and
successive mergers of lower mass stars may form massive stars (e.g.~Spitzer \& 
Saslaw 1966, Phinney 1989). 
From comparisons with samples of Galactic and LMC objects, the paucity of stars 
with similar spectra and the rare combination of low temperature and high luminosity 
observed for the GC objects has also raised the question of their evolution 
being normal (e.g.~Tamblyn \& Rieke 1993, Hanson \etal 1996, Tamblyn 1996).

\end{itemize}

First we will present a new quantitative comparison of individual GC stars
to address questions 1 and 2b (Section \ref{s_prop}). 
In a next step (Sect.~\ref{s_pop_gc}) we will review comparisons of the stellar 
population with evolutionary synthesis models, which should help to shed some 
light on question 2a.

\subsection{Properties and the nature of the \hei\ emission line objects}
\label{s_prop}
In this Section we adopt a conservative approach. From comparisons of
stellar parameters derived recently for the most luminous \hei\ 
sources with those from related objects we try to answer question 1.
We will then confront recent stellar evolution models with the observations
of individual GC stars.
If severe discrepancies between observations and our present knowledge
of stellar evolution can be found, we will presume that this may imply
that question 2b cannot be affirmed. 

Assuming the spectra of the \hei\ objects are formed in a spherically
expanding wind of a hot star Najarro \etal (1994, 1997), 
and Krabbe \etal (1995) have applied the so-called ``standard model'' of
WR stars to derive stellar parameters from fits to line profiles and the
total $K$-band flux. So far a total of nine objects have been analysed,
which we will refer to as ``the GC stars'' in the following.
The basic parameters which can be derived are the luminosity $L$, the
so-called ``core temperature'' $T_\star$ (quite strongly dependent on
specific model assumptions; cf. Schmutz \etal 1992, Schaerer 1996a) the 
relative hydrogen and helium
abundance, and the wind properties i.e.~the mass loss rate \mdot, and 
the terminal velocity \vinf.

\subsubsection{H and He abundances:}
A rough comparison shows that these parameters are in the same range of 
those derived from galactic and LMC WR stars (Hamann \etal 1995, 
Crowther \etal 1995a, Crowther \& Smith 1997). This is certainly a basic
but strong argument in favour of the massive star hypothesis.
The derived He abundances (He/H $\ge$ 1, corresponding to a hydrogen 
mass fraction $X \sim$ 0.2--0.) are {\em higher} than those in LMC Ofpe/WN9
stars (or equivalently WN9-11 according to the reclassification of Crowther
\& Smith)
which have $X \sim$ 0.3--0.5 (Crowther \& Smith 1997, Pasquali \etal 1997).
In fact the H and He abundances correspond well to the values found in
late WN stars ($\sim$ WN6-8; Crowther \etal 1995a, Hamann \etal 1995).
Abundance determinations of other elements which might bear testimony of
a more exotic nature (Morris 1993) seem hardly feasible at the present times.

\subsubsection{Observed wind momentum:}
Figure 1 (left) shows the observed radius-modified wind momentum 
$\mdot\vinf R^{0.5}$ of the GC stars (open squares) compared to Ofpe/\- WN9 
(crosses and open circles) and WNL stars (WN6-8, filled squares) in the LMC 
(Crowther \& Smith 1997, Pasquali \etal 1997)\footnote{Note that these
independent studies have 7 objects in common. Systematic differences are
discussed in Pasquali et al.}. 
As shown by Puls \etal (1996) this quantity is expected to correlate with
the luminosity $L$ if the stellar wind is driven by radiation.
As a comparison the relations followed by Galactic O stars (see Puls et al.) 
are shown by the solid and dashed lines for supergiants and LC II-V objects 
respectively. 

\begin{figure}[htb]
\centerline{\psfig{figure=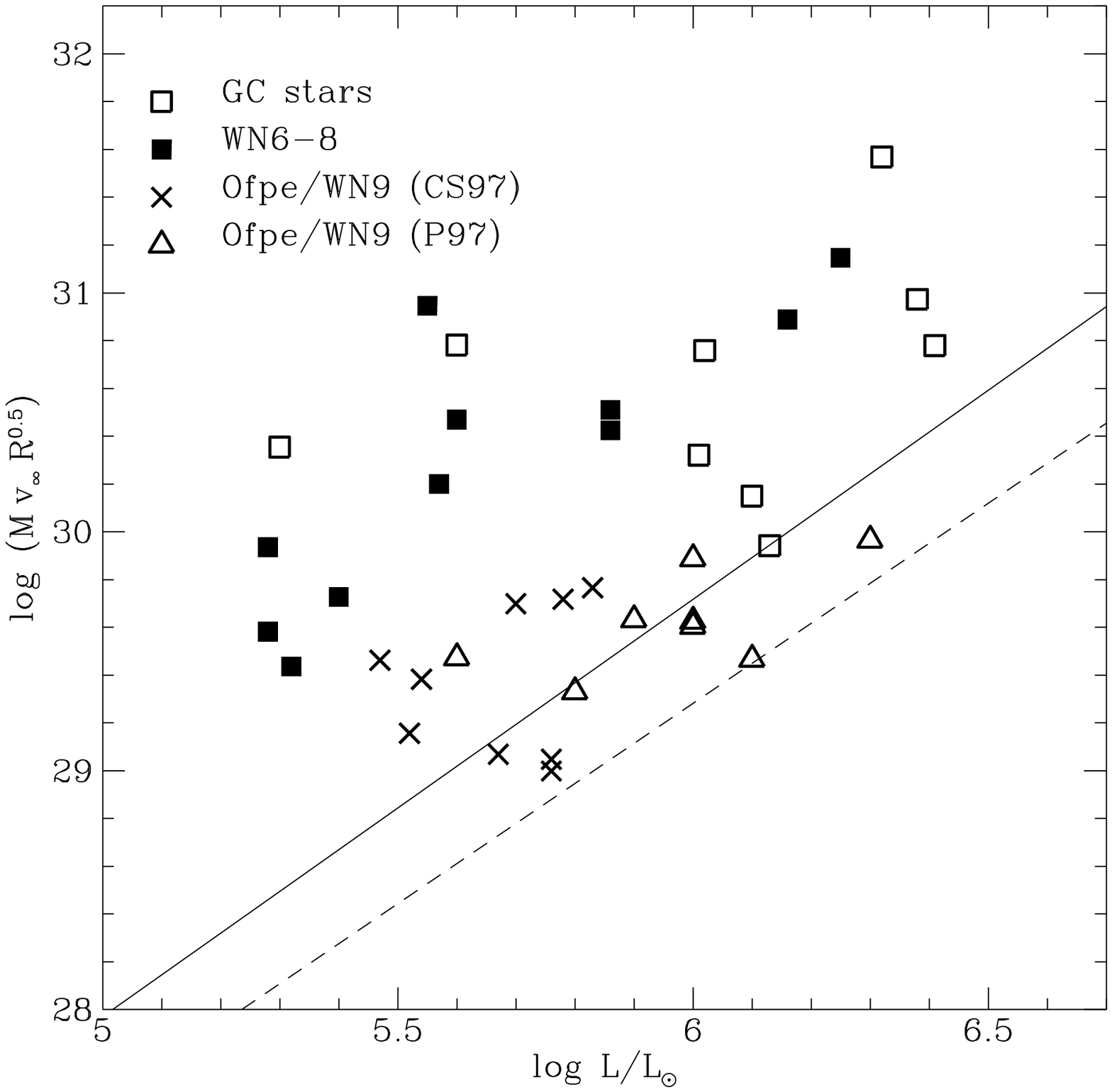,height=7cm}
	\psfig{figure=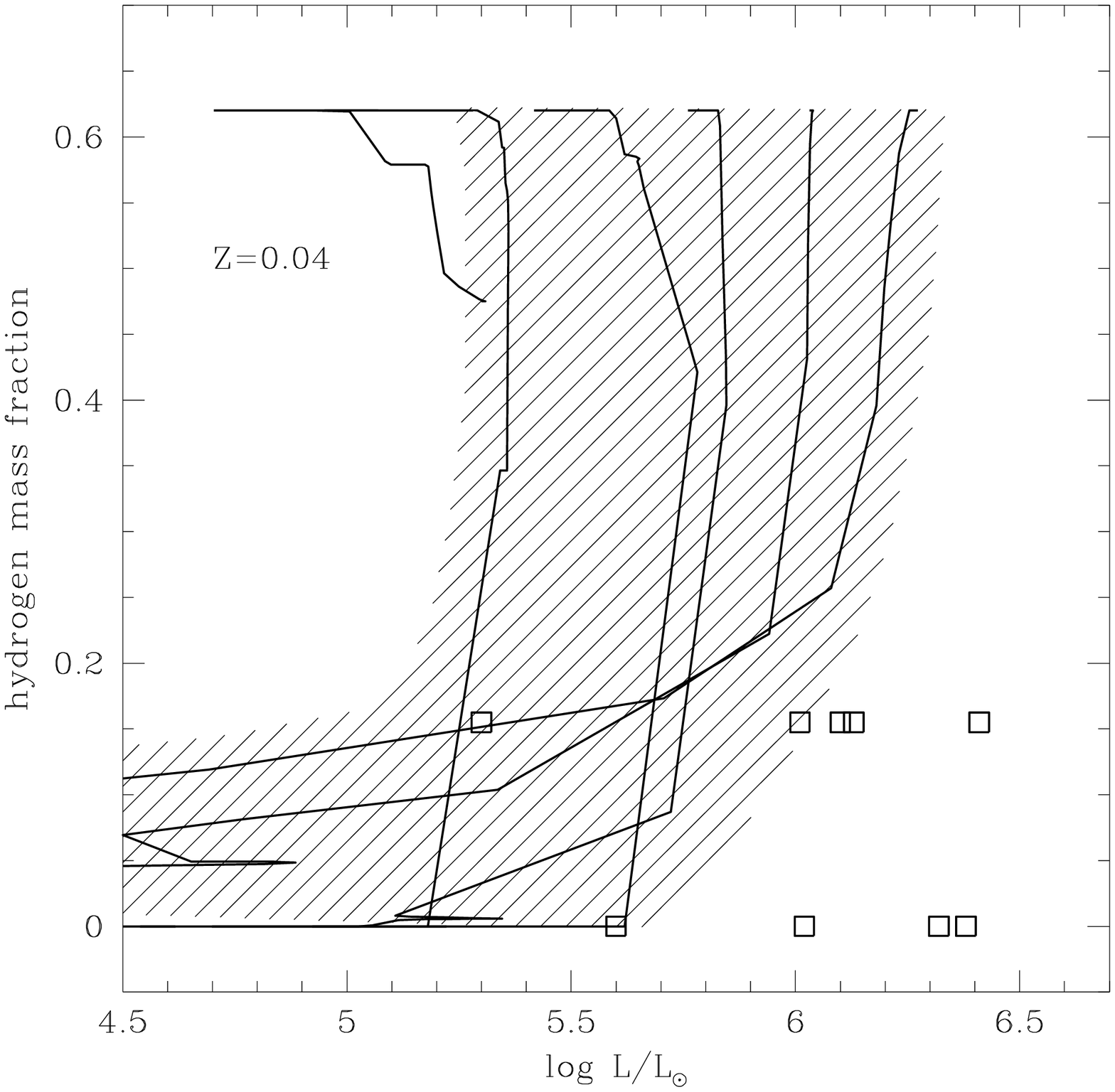,height=7cm}}
{\small
\caption{{\em Left panel:} Observed radius-modified wind momentum 
$\mdot\vinf R^{0.5}$ as a function of luminosity.
Shown are GC stars (open squares), and LMC objects: WN6-8 stars (filled
squares), and Ofpe/WN9 stars (crosses and open triangles). Sources
of the observations are given in the text.
{\em Right panel:} Comparison of predicted H-abundance versus luminosity 
from the high metallicity stellar models of Meynet \etal (1994)
with derived values for the GC stars (open squares). 
The shaded band shows the domain covered by stars with initial masses 
of $25 \le M_{\rm ini} \le 120 \msun$}
}
\label{fig_momratio}
\label{fig_lxs_gc}
\end{figure}


The slash stars follow essentially the same relation as O stars. The same also
holds for LBVs (cf.\ Leitherer 1997).
The difference between the LMC WN6-8 and the O stars is explained by the
more advanced evolutionary stage (i.e.~the higher He abundance) of the former,
which -- for yet unknown reasons -- implies ``stronger'' winds 
(see Hamann \etal 1995, Crowther \& Smith 1997). 
Interestingly enough the GC stars follow quite well the relation of the 
WN6-8 stars which may not be surprising since they share the same H/He
abundances and this quantity seems to be the determining factor for the
wind properties of WN stars.
Contrary to the winds of O stars where metallicity effects can clearly be
seen (e.g.~Kudritzki \etal 1995), such a behaviour has not been found so far
comparing WR stars between the LMC and the Galaxy (Crowther \& Smith 1997). 
We verified that the same is also true for the $L$ vs.~wind momentum ratio 
relation derived from their sample.
If the comparison between the GC stars and WNL stars is indeed appropriate
we might conclude from Fig.~1 (left) that the wind properties
of these objects do not show any metallicity effect even over a
larger metallicity range.
In fact, based on the good agreement of the {\em wind properties} of the \hei\
sources with those of WNL stars (Fig.~1), one could even 
argue that this strongly supports the hypothesis of these sources being
evolved massive stars.

\subsubsection{Comparison with evolutionary models:} 
We shall now compare the stellar parameters with predictions from 
the evolutionary models of Meynet \etal (1994) at a high metallicity
(Z=0.04) appropriate to the Galactic Center (Shields \& Ferland 1994).
Figure 1 (right) shows the evolution of the hydrogen surface
abundance as a function of the luminosity of stars with initial masses
between 25 and 120 \msun. The shaded band shows the domain covered
by all models which evolve through the WR phase. The GC stars are  
shown as open squares, the majority showing fairly high 
luminosities compared to galactic and LMC objects, except 
WN stars in R136 and NGC 3603 (de Koter \etal 1997, Crowther \& Dessart
1998). 
Due to the high mass loss, the luminosity of the most massive stars has 
already considerably decreased, when low values of $X$ are reached.
With the adopted mass loss prescription (which is however fairly uncertain,
cf.\ e.g.\ Lamers \& Cassinelli 1996, de Koter \etal 1997) 
it may be difficult to explain the most luminous GC stars even if 
an initial mass higher than 120 \msun\ is adopted. 
Uncertainties in the derivation of the stellar parameters may come 
from the model atmospheres, which do not 
include \lb\ although its effect is expected to be quite strong 
given the high metallicity and the low temperature of the GC stars 
(Schaerer 1995, Schaerer \etal 1996). On the other
hand it must be noted that the luminosities are derived from the 
$K$-band flux, which depends on extinction corrections, may be 
contaminated due to background or unresolved sources, or affected
by systematic uncertainties (cf.~Blum \etal 1996).
In view of the uncertainties in both atmosphere and evolutionary models 
we think that the luminosities can still be fairly well explained by 
the Meynet \etal evolutionary models.  
  
Figure 2 (left) shows the HR-diagram of the GC stars
and the Meynet \etal (1994) evolutionary tracks (wind corrected 
\teff\ from the tracks).
Plotted are the ``core temperatures'' $T_\star$, which are typically
1000 to 5000 K larger than the ``photospheric'' values (cf.~Najarro 1995).
It is well known that for WR stars such comparisons are 
hampered by the lack of understanding the hydrodynamics of their 
stellar winds (see e.g.~Schaerer 1996a). The comparison may 
therefore only be indicative. Interestingly, however, most of the
GC stars are found in a relatively narrow temperature range 
($\log T \sim$ 4.3 -- 4.4) which coincides with the domain
populated by the hydrogen burning WNL stars descending from the
most massive stars ($M_{\rm ini} >$ 40 \msun).
 
\begin{figure}[ht]
\centerline{\psfig{figure=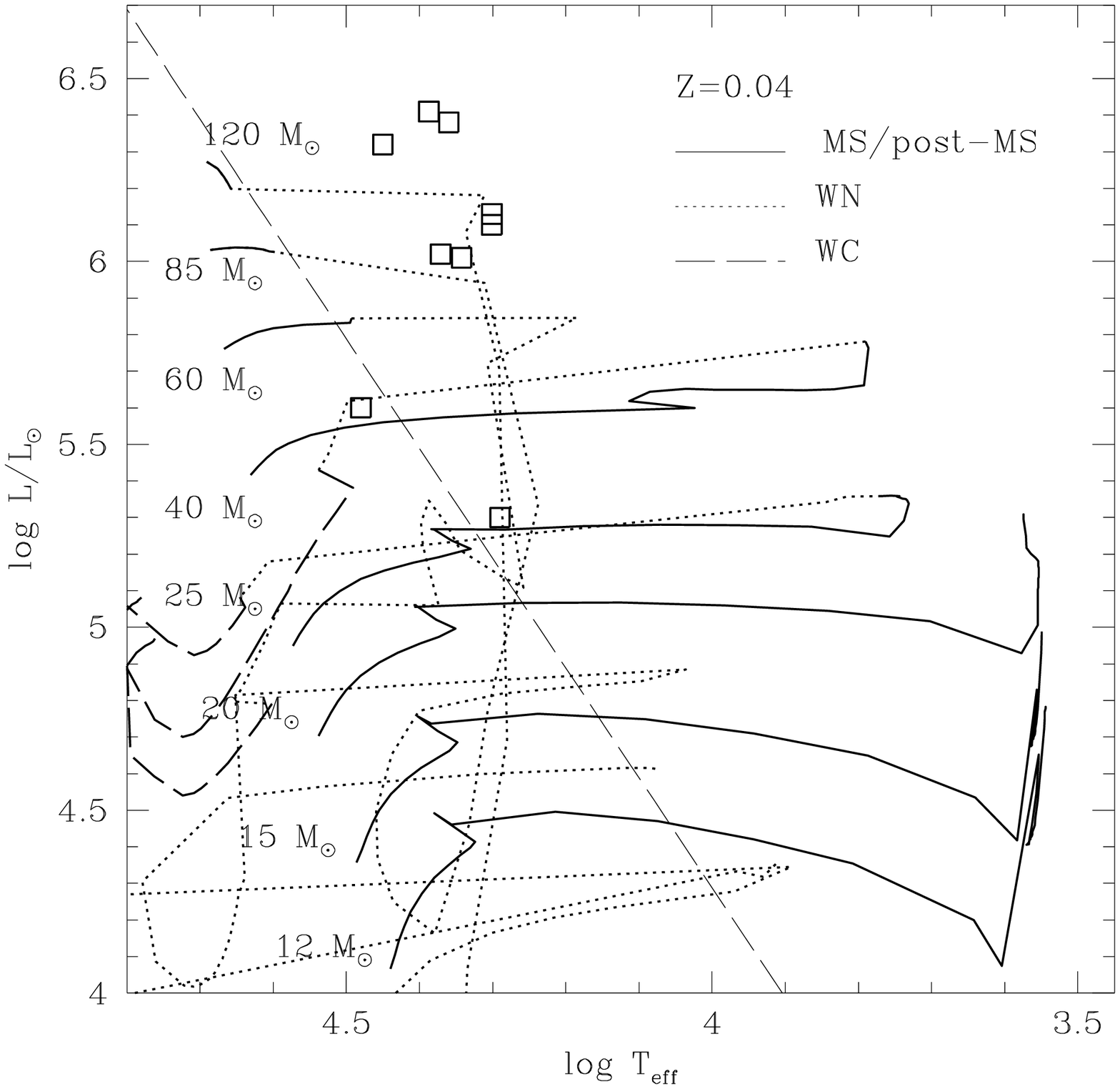,height=7cm}
	\psfig{figure=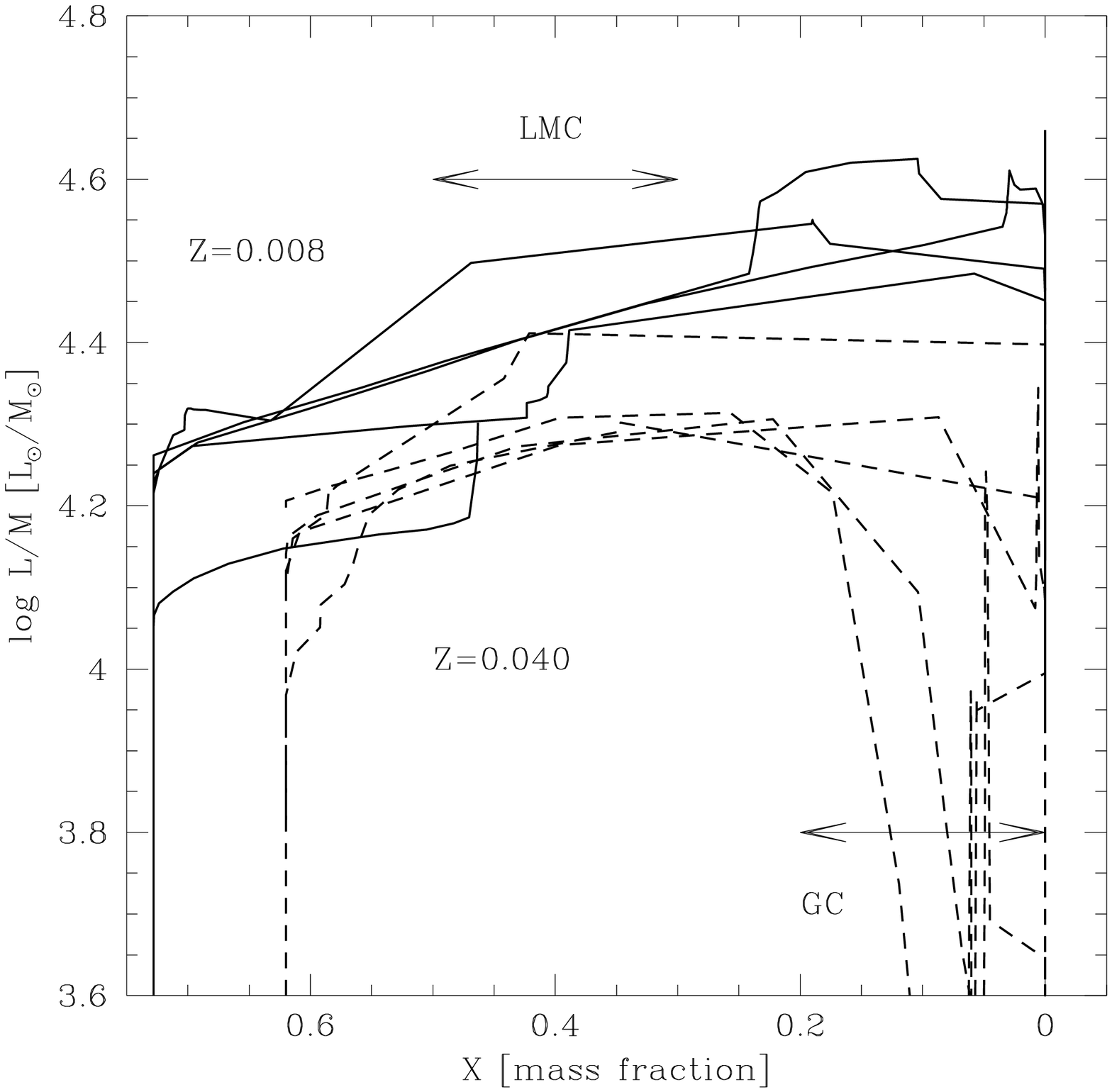,height=7cm}}
{\small
\caption{{\em Left panel:} HR-diagram comparing the GC stars (open squares)
with $Z=0.04$ evolutionary tracks from Meynet et al. Solid lines indicate
main-sequence and post-MS phases, dotted lines the WN phase 
(both H or He-burning), and long-dashed lines WC/WO phases. 
The dashed diagonal line shows an estimate of the background limit at 
$m_K=11.8$ (see text).
{\em Right panel:} Logarithm of the luminosity to mass ratio as a function
of the H surface abundance predicted for the LMC ($Z=0.008$ tracks, solid 
lines) and the GC ($Z=0.04$, dashed lines). The different lines correspond
to the values from the different individual stellar tracks of Meynet
et al.
The arrows indicate the range of the observed H abundances in LMC Ofpe/WN
stars and the GC stars. The systematic differences of $L/M$ between 
low and high $Z$ tracks might explain differences in wind velocities.
See discussion in the text}
}
\label{fig_hrd_gc}
\label{fig_l_m}
\end{figure}


Overplotted on Fig.~2 (diagonal dashed line) is an 
estimate of the background limit at $m_K=$ 11.8 following Tamblyn 
(1996) assuming a black-body spectrum, the extinction law 
of Rieke \& Lebofsky (1985), and an extinction at $K$ equivalent
to $A_V=$ 30.
This shows that (apart from red supergiants) most of the objects 
above this limit are expected
to be hydrogen- and helium burning WN stars, while O stars above 
the background limit should only be found in a relatively narrow 
luminosity range ($\log L/\lsun \sim$ 5.5--5.7).
This seems to be consistent with the present lack of observed 
O stars (cf.~Genzel \etal 1994).
The indicated limit is not in conflict with the detection of 
evolved WC stars (Blum \etal 1995a, Krabbe \etal 1995) since their 
bolometric correction differs considerably from the assumed 
black-body value (see e.g.~Blum \etal 1995a). Indeed the 
finding of WC stars of spectral type WC9 agrees well with the
expectations from evolutionary models, which explain why late
WC stars should only be found in high metallicity environments
(cf.~Maeder 1991a, Maeder \& Meynet 1994).

\subsubsection{Unusual wind velocities ?}
As mentioned earlier, the GC stars show larger terminal velocities
than the Ofpe/WN9 stars in the LMC, the average value being
a factor of 2.3 larger (cf.~Najarro 1995 and Crowther \& Smith 
1997; see also Blum \etal 1995b). 
Due to the metallicity difference one would expect only a modest 
increase of \vinf\ ($\sim$ 30 \%) from the radiation driven 
wind theory (Kudritzki \etal 1987, Leitherer \etal 1992).
From a comparison of evolutionary models it appears
that another largely unnoted systematic difference between 
evolved stars at different metallicities may, however, explain
such changes more easily as we will show in the following.

Given the lower initial H abundance and the large mass loss rates,
low surface H abundances corresponding to Ofpe/WN and WR stars are 
attained more rapidly at high metallicity than at low $Z$.
At a given H surface abundance the interior of WN stars will thus,
on the average, be less evolved at high metallicity. This in 
particular implies that the luminosity to mass ratio $L/M$ is 
smaller in high $Z$ models for a given surface abundance $X$.
Figure 2 (right) illustrates this behaviour by comparing
the $L/M$ ratio from low metallicity tracks appropriate 
to LMC composition (solid lines, $Z=$0.008, Meynet et al.) 
and high $Z$ models (dashed lines). Also shown is the abundance 
range determined for Ofpe/WN9 stars in the LMC and the GC stars.   

The difference in $L/M$ may have the following bearing:
Since the ratio $\Gamma=L/L_{\rm Edd}=\kappa L /(4\pi G c M)$ of 
the luminosity to the Eddington luminosity is proportional to
$L/M$, for a given opacity $\kappa$ lower $Z$ models have on 
the average larger values of $\Gamma$ in evolved stars where H 
is still present.
If we assume that we can at least qualitatively apply the radiation 
driven wind theory to these stars one expects
\[
\vinf^2 = \frac{\alpha}{1-\alpha}\frac{2 G M}{R}(1-\Gamma)
\]
(Castor \etal 1975), i.e.~the closer proximity to the 
Eddington limit implies {\em lower terminal wind velocities} 
for the lower metallicity models. 
We suggest that this systematic difference may, at least 
qualitatively, explain the larger observed terminal velocities 
of the GC stars compared to the Ofpe/WN9 objects in the LMC. 
A more rigorous quantitative understanding would not only require 
to account simultaneously for differences in metallicity and H/He 
composition but also for the apparent temperature differences
which are not fully understood yet (cf.~below).

\subsubsection{Unusual temperatures ?}
The most puzzling feature of the GC stars seems to be their 
temperature which is lower than that of LMC and 
Galactic Ofpe/WN9 and WNL stars.
The indication of low temperatures is primarily supported
by the predominance of \hei\ and the weakness or even absence of 
\heii\ lines in the $K$ band spectra of most objects (Blum \etal 
1995b, Libonate \etal 1995). 
The core temperatures $T_\star$ (``photospheric'' temperatures 
$T_{2/3}$) of all GC stars analysed by Najarro (1995) and 
Krabbe \etal (1995) are in the range of
$T_\star \sim$ 20 -- 30.4 ($T_{2/3} \sim$ 18.8 -- 28.9) kK, compared to 
31.2 -- 35.9 (24.7 -- 32.5) kK for Galactic WNL, and 
27.9 -- 39.4 (25.4 -- 32.9) kK for LMC WNL stars 
(Crowther \etal 1995a, Crowther \& Smith 1997).
Typically both temperatures are lower by 0.1 dex in the GC stars
compared to Galactic and LMC Ofpe/WN9 and WNL stars, while between the
latter no significant difference is apparent.

Contrary to claims of Tamblyn \etal (1996) and Tamblyn (1996)
we will now argue that a large metallicity in the GC may well
play a role explaining the above differences for the following reasons:
{\em 1)} Line blanketing is not included in the atmosphere models 
used in the analysis. 
As pointed out by Schaerer (1995) and Schaerer \etal (1996) blanketing
is expected to be of particular importance for objects similar to the 
AF star and should hence be included in future spectroscopic analysis.
%
{\em 2)} The wind properties of the GC sources might differ as would be 
expected if the driving mechanism of WR wind is closely related to the 
iron opacity peak (Schaerer \etal 1995, Pistinner \& Eichler 1995).
So far the comparison of wind properties (see above) does, however, not 
reveal any significant difference.
{\em 3)} The feedback mechanism between strong wind blanketing and
a thin subphotospheric convection zone pointed out by Schaerer 
(1996a) may maintain a larger radius.
{\em 4)} Last, but not least, the temperature differences between
the GC stars and Galactic/LMC WNL stars might also simply be due 
to differences in their evolutionary status (majority core H-burning
versus core He-burning objects).

Given our poor knowledge of the winds of Ofpe/WN and WR stars, the
difficulties in deriving temperatures and radii of WR stars 
(see e.g.~Moffat \& Marchenko 1996, Schmutz 1997, Schaerer 1996a),
and the uncertainties mentioned above,
we presently do not consider the temperatures of the GC stars as a 
strong constraint on their nature and/or evolution.
Future progress on this issue would, however, be extremely interesting.

\subsubsection{Summary:}
From previous investigations and the properties discussed above 
it can be concluded that the emission line objects in the GC cluster
are compatible with massive evolved stars. They share the surface 
abundances and wind properties of WNL stars rather than those of 
Ofpe/WN9 stars to which
they are mostly associated based on their $K$-band spectra.
Their properties are in fair agreement with predictions from standard 
evolutionary models at high metallicities, which indicate that the GC
stars can be both H or He burning objects.
Within the remaining observational and theoretical uncertainties
there is no compelling evidence that the individual stars have undergone 
an unusual evolution.

We note, however, that the results derived from comparisons with evolutionary 
models rely quite strongly on the large adopted mass loss rates \mdot\ 
thought to be representative for the high metallicity in the GC 
(cf.~Shields \& Ferland 1994). However, interestingly Carr \etal (1996) 
derived a roughly solar metallicity for the M2 supergiant IRS 7.
The presence of additional mixing processes (cf.~Sect.~\ref{s_evol}) 
can to a certain extent have similar effects than large mass loss 
rates. The implications of such alternate evolutionary models will 
be considered in the future.

\subsection{The Quintuplet and the Arches cluster}
\label{s_arches}
In addition to the central cluster two more spectacular clusters of
young stars are now known: the Quintuplet cluster ($=$ AFGL 2004) and
the G0.121+0.017 ($=$ Object 17, or ``Arches cluster'', hereafter used), 
both located approximately within 30 pc projected distance of the GC
(see review by Morris \& Serabyn 1996 and references therein).
After the discovery of emission line stars in the Quintuplet and 
the Arches cluster (cf.\ Nagata \etal 1990, 1995, Cotera \etal 1994, 1996) 
a great wealth of new data has been obtained very recently about these 
clusters.

The observations of Figer \etal (1996, 1998a) of the Quintuplet reveal 
a cluster with $\sim$ 8 WR stars and approximately a dozen other stars in
earlier stages of evolution. Probably associated with it is the so-called
``Pistol'' star (Figer \etal 1998b, an LBV candidate of very 
high luminosity if single.
13 emission line stars have been identified in the Arches cluster by 
Cotera \etal (1996). If all emission line stars in these clusters are
WR stars they represent an important increase of the known Galactic
WR population (van der Hucht 1996).
The mere finding of additional emission line stars similar to the
one in the central cluster has also been taken as argument against
their being exceptional (Cotera \etal 1996, Figer \etal 1998a).

In the recent Keck images of the Arches by Serabyn \etal (1998) massive
main-sequence stars (probably OB types) have quite likely been detected 
for the first time in one of the GC clusters.
The masses of the Quintuplet and the Arches are of the order of 1000 --
5000 \msun\ respectively for the observed stars; extrapolation of a Salpeter
IMF down 1 \msun\ yields masses larger by factors 4-6.
The Arches cluster is of similar compactness and stellar density as the 
central cluster; the estimated stellar density is $\sim$ 2 order of magnitude
lower in the Quintuplet (Figer \etal 1998a).

The richness in massive stars and the diversity of densities make the
three GC clusters an exceptional field for studies of massive stars in different
environments. While quantitative work has been done on massive stars in the
central cluster all remaining objects (except the Pistol star, Figer \etal
1998b) await future analysis. 
New upcoming high angular resolution IR observations ({\it HST} NICMOS, 
adaptive optics work etc.) will also provide a wide field of investigation.

Let us now go back and briefly review the status of the stellar population
of the central cluster as a whole.

\subsection{The stellar population in the central cluster}
\label{s_pop_gc}
A fair number of observational constraints (number of stars of different 
types, total mass, $L_{\rm Lyc}/L_{\rm bol}$ etc.) clearly show that the GC
cluster cannot be explained by a constant star formation rate and a 
standard IMF (e.g.~Genzel \etal 1994).
Burst models adapted to the GC stars have therefore been studied
by Tamblyn \& Rieke (1993), Krabbe \etal (1995), Schaerer (1996b), and
Tamblyn (1996) and have led to differing conclusions.
Although all these studies agree on the fact that recent 
star formation is required to describe the massive star population, Tamblyn (1996) 
argues that high luminosities and the spectroscopic ``uniqueness'' of the \hei\ stars 
show that they cannot arise in such numbers from normal stellar evolution.

If the luminosity spread of the evolved WR like objects is as large
as shown in Fig.~2 (left) and their evolution is ``normal'' they cannot 
be coeval.
Indeed, their ages are between $\sim$ 1.5 and 5 Myr based on isochrone fitting  
in the HR diagram - an age spread which is roughly in agreement with that 
observed in young clusters (Massey \etal 1995 and Sect.~\ref{s_clusters}). 
Krabbe \etal (1995) find that a decaying burst beginning $\sim$ 7 Myr
ago and a decay time of 3--4 Myr can explain both the population of early and
late type stars very well and that the hot star cluster can well account for 
the bolometric and ionizing luminosities of the central parsec.
However, one has to remember that the observed number of
stars is still small and that less evolved main sequence stars have not been
detected yet (cf.~Genzel \etal 1994).
In our opinion the findings summarized in this paragraph show that burst models 
do not allow us to identify any signs of ``unusual'' stellar evolution
of massive stars if the history of recent star formation cannot be known better.

We conclude with a remark about less massive stars, which have not been 
discussed in this study:
Late type stars representing an important population
in the GC, show very distinct properties (see Blum \etal 1996, Genzel \etal 1996). 
In particular they may well indicate
that collisions between red giants and MS stars might have occurred in the
dense stellar core (see e.g.~Sellgren \etal 1990, Genzel \etal 1994, 1996).
The Galactic Center is a rich and fascinating field whith many plots 
still to be unraveled !




\acknowledgments
It is a pleasure to thank the organizers and hosts of this 
workshop, Judith Perry, Susan Lamb, and Jayant Narlikar 
for their work and their hospitality which greatly contributed to the 
``multidisciplinary'' spirit of this meeting.
I also thank Bob Blum, Claus Leitherer, Andr\'e Maeder, and Nolan
Walborn for useful discussions and comments.

This work was supported by the Swiss National Foundation of Scientific 
Research. Additional support from the Directors
Discretionary Research Fund of the STScI and a travel grant from
the NSF is also acknowledged.
Finally I thank the Institute for Nuclear Theory at the University
of Washington for its hospitality and the Department of Energy for
partial support during the writing of the first version (july 1996) 
of this work.

\end{document}